\documentclass[prl,twoside,preprintnumbers,superscriptaddress,twocolumn,nofootinbib]{revtex4}

\usepackage{amsmath,graphicx,slashed,multirow,color,ulem}
\usepackage{graphicx,epsfig,amssymb,bm,slashed,wasysym,multirow}
\def \beq{\begin{equation}}
\def \eeq{\end{equation}}
\def \beqa{\begin{eqnarray}}
\def \eeqa{\end{eqnarray}}

\usepackage{mciteplus}
\begin{document}

\preprint{DCPT/15/110}
\preprint{IPPP/15/55}

\title{Forward $D$ predictions for $p\rm Pb$ collisions, and sensitivity to cold nuclear matter effects}

\author{Rhorry Gauld}
\email{rhorry.gauld@durham.ac.uk}
\affiliation{Institute for Particle Physics Phenomenology, University of Durham, DH1 3LE Durham, United Kingdom}

\date{\today}


\begin{abstract} 
Predictions are provided for double differential cross sections and forward-backward ratios of $D^{0}$ production in $p\rm Pb$ (forward) and $\rm{Pb}$$p$ (backward) collisions at 5.02~TeV. The effect of nuclear corrections on the ratio of differential cross sections ratios is estimated to be $\simeq$ (10-30)\% in the kinematically accessible region of LHCb, and interestingly this ratio is approximately flat with respect to $p_T(D^0)$ due to a compensation of shadowing and anti-shadowing effects arising from the input nuclear PDFs.  In comparison to $J/\psi(\rightarrow\mu^-\mu^+)$ measurements which have already been performed with the available data, the cross section for $D^{0}(\rightarrow K^-\pi^+)$ production is expected to be two-orders of magnitude higher.
\end{abstract}

\maketitle

\section{Introduction}

Measurements of $D$ hadron production in Pb-Pb~\cite{ALICE:2012ab} and Au-Au~\cite{Abelev:2014hha} collisions show evidence for the suppression of the differential $D$ cross section as compared to references $pp$ collisions. This suppression can be successfully described by in-medium energy loss effects expected in the presence of a Quark-Gluon Plasma~\cite{Uphoff:2012gb,Wicks:2005gt,Horowitz:2011wm,Sharma:2009hn,He:2014cla}.
However, a suppression of $D$ production in heavy ion collisions is also expected in the absence of a hot nuclear medium due to cold nuclear matter (CNM) effects~\cite{Vitev:2006bi,Vitev:2007ve,Neufeld:2010dz,Kang:2015mta,Andronic:2015wma} alone. 
These effects, such as the coherent, incoherent and inelastic scattering in nuclear matter, as well as a modification of the effective parton luminosities, arise as the colliding constituent nucleons of the heavy ion are not free. 
To interpret the suppression of $D$ production in heavy ion collisions due to hot medium effects, it is necessary to first quantify the size of CNM effects with independent measurements. One way of disentangling these effects is to perform measurements in $pA$ collisions, where only CNM effects are expected to be present.

Measurements of $J/\psi$ production in $p\rm{Pb}$ collisions at 5.02~TeV, which are subject to these CNM effects, have been performed by both ALICE and LHCb collaborations~\cite{Abelev:2013yxa,Aaij:2013zxa}. In both cases, differential measurements of the ratio of $J/\psi$ production in $p\rm Pb$ (forward) and $\rm{Pb}$$p$ (backward) collisions in the nucleon-nucleon centre-of-mass (COM) frame are presented, an observable defined as
\beq \label{Rfb}
\begin{aligned}
\frac{ d R_{fb}}{dx} \equiv \frac{d\sigma^{pPb}(x)}{dx}\Bigg/\frac{d\sigma^{Pbp}(x)}{dx}\, .
\end{aligned}
\eeq
For the case of the above mentioned $J/\psi$ measurements the observable $x$ denotes transverse momentum ($p_T$) and rapidity ($y$), and a sizeable suppression is observed by both collaborations. The data has also been compared to LO and NLO predictions which incorporate CNM effects through a nuclear modification of the free proton and neutron parton distribution functions (PDFs). The NLO predictions~\cite{Arleo:2013zua,Vogt:2015uba}, which describe both $p_T$ and $y$ distributions, incorporate the NLO EPS09 nuclear PDF (nPDF) modifications~\cite{Eskola:2009uj} and provide a reasonable description of the data --- in particular the calculation based on a parton energy loss model~\cite{Arleo:2013zua}. 
Although this indicates that the EPS09 nPDFs describe the dominant CNM effect, it is clear that either additional effects such as parton energy loss or altered heavy-quark fragmentation functions~\cite{Vogt:2015zoa} or refitted nPDFs with more low-$x$ data (which may in fact themselves be parameterising such effects) are required to describe the observed $J/\psi$ data.
Given that the size of CNM effects in $p\rm Pb$ lead collisions are important for the interpretation of the observed $D$ (and $J/\psi$~\cite{Abelev:2013ila}) suppression in Pb-Pb collisions, they should be validated with measurements of other final states in a similar kinematic regime\footnote{The ALICE collaboration has presented a measurement of the rate of $D$ production in $p\rm{Pb}$ collisions in the central region~\cite{Abelev:2014hha}. As compared to a $pp$ reference, these results are consistent with unity within large uncertainties of about (15-20)\%.}.

It has previously been shown that pQCD predictions~\cite{Kniehl:2004fy,Kniehl:2005de,Kneesch:2007ey,Kniehl:2009ar,Kniehl:2012ti,Cacciari:1998it,Cacciari:2005uk,Cacciari:2012ny,Zenaiev:2015rfa,Gauld:2015yia} provide a satisfactory description of the forward $D$ production as presented by the LHCb collaboration for 7~TeV $pp$ collisions~\cite{Aaij:2013mga}. In recent work~\cite{Zenaiev:2015rfa,Gauld:2015yia}, it then been demonstrated how the inclusion of this data in a global QCD analysis of the proton provides substantial improvement in the description of the low-$x$ gluon PDF. The main point being that, forward $D$ production at low-$p_T$ provides sensitivity to incoming partons at moderate (low) values of Bjoerken-$x_{1,(2)} \simeq 2\cdot10^{-2}(5\cdot10^{-5})$, where $x_{1,(2)}$ is the fraction of momentum carried by the constituent parton of the forward (backward) travelling proton. Therefore, measurements of $D$ production probe a similar kinematic regime to $J/\psi$ production, albeit with slightly larger average values for $x_1$, $x_2$, and $Q^2$. 

As the ratio of $D^0(\rightarrow K^-\pi^+)$~\cite{Aaij:2013mga} ($D^0$ will refer to the sum of $D^0$ and $\overline{D}^0$ mesons) and $J/\psi(\rightarrow \mu^-\mu^+)$~\cite{Aaij:2011jh} production cross sections measured within the LHCb fiducial region in $pp$ collisions at 7~TeV is approximately $100$, and even larger for moderate $p_T$ values, a significant improvement in the statistical precision of differential $R_{fb}$ measurements in $p\rm Pb$ collisions can be expected for $D$ hadrons as compared to $J/\psi$. Furthermore, the relative systematic uncertainty of the double differential $D^0$ measurements performed in $pp$ collisions is slightly smaller than those for the corresponding $J/\psi$ measurement. One therefore naively expects that differential $R_{fb}$ measurements of $D$ hadrons will have improved precision as compared to the $J/\psi$~\cite{Aaij:2013zxa} counter-part.

In this letter NLO+LL predictions are provided for double differential $D$ production within the LHCb fiducial region for both forward and backward $p\rm Pb$ collisions at 5.02~TeV. Throughout, CNM effects are incorporated via EPS09 nPDF modifications~\cite{Eskola:2009uj}. Finally, differential predictions for the observable $R_{fb}$ are also provided.

\section{Heavy-Quark Preliminaries}

The centrality integrated cross-section for heavy-quark pair production in $pA$ collisions can be computed by applying the standard factorisation formula as
\beq
\begin{aligned}
\sigma_{pA\rightarrow Q\bar{Q}} = A \sum_{i,j} &\int dx_{i,j}  f_{i/p}(x_i,\mu_F^2) R^{\rm{nuc}}_{j/A} \otimes f_{j/N}(x_j,\mu_F^2) \\
&d\hat{\sigma}_{ij\rightarrow Q\bar{Q}X}\left(\hat{s},\mu_F^2,\mu_R^2,\alpha_s(\mu_R^2),m_Q\right) + ...
\end{aligned}
\eeq
%
where the ellipses denote non-factorisable corrections which are neglected in this approximation. This differs from the corresponding cross section in $pp$ collisions by the introduction of the term $R^{\rm{nuc}}_{j/A} \otimes f_{j/B}(x_j,\mu_F^2)$, which represents a flavour and mass number ($A$) dependent nuclear modification factor to the parton distribution functions of the colliding nucleon, and a linear scaling of the cross section with $A$.

Throughout this work, the EPS09 nPDF modifications are adopted as a baseline. These modification parameters, which are applied to free proton or neutron PDFs, are provided in a Hessian basis which allows the uncertainties from the global EPS09 analysis to be propagated to observables. The colliding Pb atom will also be approximated to be wholly constituted of protons, which is reasonable for the QCD production of heavy-quarks which is insensitive to the flavour of high-$x$ valence quarks.

It is worth commenting that although the EPS09 nPDFs have been extracted using CTEQ6.1M PDFs~\cite{Stump:2003yu}, the application of these nPDFs to other input proton PDFs has been shown to have little to no impact on $p\rm Pb$ to $pp$ cross section ratio observables such as $R_{p\rm Pb}$~\cite{Vogt:2015uba}. Checks with the LO nPDF modifications applied to NNPDF3.0 LO~\cite{Ball:2014uwa} and cteq6l1~\cite{Pumplin:2002vw} proton PDFs indicate this is also true for $R_{fb}$. As a baseline PDF set, the EPS09 nPDFs are applied to the proton VFNS NNPDF3.0LHCb NLO PDF set with $\alpha_s(m_Z)$ =~0.118 from~\cite{Gauld:2015yia}. This PDF set was obtained by including 7~TeV LHCb $D^{\pm}$ and $D^{0}$ data~\cite{Aaij:2013mga} into the NNPDF3.0 global data set via the standard Bayesian reweighting technique~\cite{Ball:2010gb,Ball:2011gg}. This PDF set has the advantage of an improved description of the behaviour of the gluon PDF at low-$x$, and is also provided in the same scheme as the nPDFs. All PDFs are accessed via the LHAPDF6 interface~\cite{Buckley:2014ana}.

The heavy-quark predictions are computed at NLO using the massive calculation~\cite{Nason:1989zy} implemented in {\sc\small POWHEG-BOX}~\cite{Frixione:2007nw}. This calculation is performed in a fixed-flavour scheme (FFS), where the heavy-quark flavour is considered massive, and not included as a degree of freedom in the running of $\alpha_s$ or in the PDFs. To consistently convolute this calculation with the five-flavour PDFs and nuclear corrections, it is necessary to perform a change of renormalisation scheme as explained in~\cite{Cacciari:1998it}. For charm production, it is necessary to add the following corrections factors in the region $\mu_{F,R}^2 > m_b^2$
\beq
\begin{aligned}
-\hat{\sigma}^{(0)}_{q\bar{q}} 	& \frac{2 T_F \alpha_s(\mu_R^2)}{3\pi} \left( \rm{Log}\left[ \frac{\mu_R^2}{m_c^2}\right] + \rm{Log}\left[ \frac{\mu_R^2}{m_b^2}\right] \right) \,,\\
-\hat{\sigma}^{(0)}_{gg} 		& \frac{4 T_F \alpha_s(\mu_R^2)}{3\pi} \rm{Log}\left[ \frac{\mu_R^2}{\mu_F^2}\right] \,,
\end{aligned}
\eeq
where $\hat{\sigma}^{(0)}_{ij}$ represents the Born contribution. The first term accounts for the change in the running of $\alpha_s$ with $n_f+2$ active flavours, and the second term accounts for both $\alpha_s$ modifications and the depletion of the gluon PDF due to $g\rightarrow Q\bar{Q}$ splittings above charm and bottom quark mass thresholds. In the kinematic region $m_c < \mu^2 < m_b^2$, only the correction factors for one flavour are necessary. Predictions provided in this way are consistent with those obtained in a FFS to sub-leading terms of $\mathcal{O}(\alpha_s^4)$\footnote{These corrections were already implemented for similar predictions provided in other work~\cite{Gauld:2015qha}.}.
This calculation is then subsequently interfaced to {\sc\small Pythia8}~\cite{Sjostrand:2007gs,Sjostrand:2014zea} using the {\sc\small POWHEG} method~\cite{Nason:2004rx,Frixione:2007vw,Alioli:2010xd}, and the default {\sc\small Pythia8} {\sc\small Monash 2013} tune~\cite{Skands:2014pea} is used throughout. The hadronisation of heavy-quarks in this set-up is described by a non-perturbative model for which the modelling parameters have been tuned to LEP data. The predictions of $D$ hadrons in this work is therefore accurate at NLO+LL accuracy, due to the collinear resummation in the parton shower.

When providing predictions for observables, the following sources of theoretical uncertainty will be considered: charm mass, nPDFs, proton PDFs, and missing higher-order corrections. The charm quark pole mass is taken to be $m_c = (1.5\pm0.2)$~GeV and the corresponding uncertainty is found by taking the envelope of predictions found with the variation $\delta m_c = 0.2$~GeV. The nPDF uncertainty is found by computing the asymmetric 1$\sigma$ CL from the eigenvectors basis, after scaling the eigenvector deviations down by 1.645. The PDF uncertainty is found by computing the 1$\sigma$ CL from the replica set. A scale uncertainty, due to missing higher-order corrections, is evaluated by varying factorisation and renormalisation scales independently by a factor of two around the nominal scale $\mu$ with the constraint $1/2 < \mu_F/\mu_R < 2$. The nominal scale is set as $\mu = \sqrt{m_c^2 + p_T^2}$, where $p_T$ is the heavy-quark transverse momentum in the underlying Born configuration. The sum of these contributions added in quadrature is taken to be the total uncertainty.

Finally, it is worth noting that in very low-$p_T$ region ($< 1$~GeV), extremely low scales are probed when performing scale variation. 
The baseline NNPDF3.0LHCb PDFs are only provided for scales above $\mu_F > 1$~GeV, and the nPDFs $R^{\rm{nuc}}_{i/A} (x_i,\mu_F)$ are only valid in the region ($x > 10^{-4}$, $\mu_F > 1.3$~GeV), below which the (n)PDFs are frozen. For this reason, a lower $p_T$ cut of 1~GeV is enforced on observables which are $p_T$-integrated, which reduces the dependence of the observable on the poorly described low-$Q^2$ region. An alternative approach would be to perform an extrapolation of (n)PDFs, which is now available for PDFs by default as of LHAPDF 6.1.5, and take the envelope of the two approaches as an uncertainty.

\section{Cross section predictions and kinematics}
During the $p\rm Pb$ run of the LHC in early 2013, LHCb recorded data samples corresponding to integrated luminosities of 1.1~nb$^{-1}$ and 0.5~nb$^{-1}$ in forward and backward collisions respectively. The energy of the proton beam was measured to be $E_p =$ 3988~GeV, while that of the Pb beam is $Z/A \cdot E_p =$ 1572~GeV per nucleon, resulting in a proton-nucleon COM energy of $\sqrt{s_{pn}} =$ 5008~GeV~\cite{Aaij:2014pvu} (referred to as 5~TeV). Due to the asymmetry of the proton and nucleon beam energies in the lab frame, there is a rapidity shift between the COM and lab frames of $\Delta y =$ +0.465 for forward collisions and $\Delta y =$ -0.465 for backward collisions. As the LHCb detector is only instrumented in the forward region $2.0 < \eta < 4.5$, this results in an effective coverage of $1.53 < y < 4.03$ and $-2.47 < y < -4.97$ for forward $p\rm Pb$ collisions. For data taking during $p\rm Pb$ collisions in Run-II, predictions for a proton beam energy of $E_p =$ 6500~GeV, and corresponding Pb beam energy of $Z/A \cdot E_p =$ 2563~GeV per nucleon, corresponding to a proton-nucleon COM energy of $\sqrt{s_{pN}} =$ 8163~GeV have also been computed. These predictions are provided in the Appendix.

\begin{figure}[ht!]
\begin{center}
\makebox{\includegraphics[width=1.1\columnwidth]{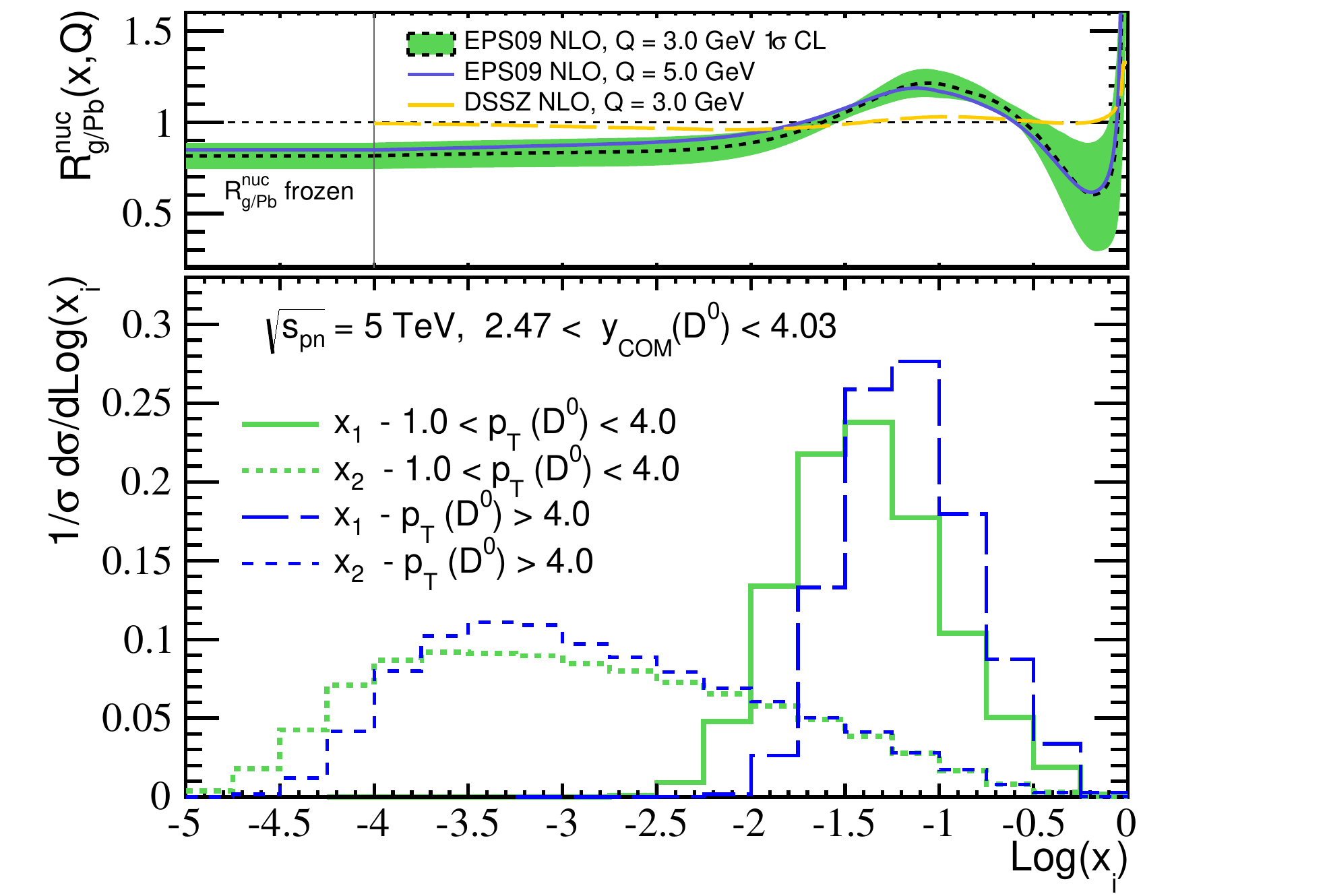}}
\end{center}
\vspace{-8mm}
\caption{Upper region: NLO EPS09 gluon nuclear modification factor $R^{\rm{nuc}}_{g/Pb} (x_g,Q)$ for $Q = 3, 5$~GeV. Lower region: Normalised differential $D^0$ cross section with respect to $\rm{Log}$$(x_i)$. The $D^0$ is required to be within the COM overlap region of forward and backward collisions accessible to LHCb, and the effect of placing $p_T$ cuts on the $D^0$ is indicated.}
\label{fig:Kin}
\end{figure}

Before providing double differential cross sections for $D$ production, it is informative to first study how the forward acceptance and $p_T$ cuts effect the sampling of the nPDFs. In the lower region of Figure~\ref{fig:Kin}, the normalised differential $D^0$ cross section with respect $\rm{Log}$$(x_i)$ at 5~TeV is shown for low- and high-$p_T$ regions. In both cases, the $D^0$ rapidity is limited to the COM overlap region of forward and backward collisions ($2.47 < y_{\rm COM} < 4.03$). The blue lines indicate the sampling when a minimum $p_T$ cut of 4~GeV is required, which shifts both $x_1$ and $x_2$ to larger values as compared to the low-$p_T$ region --- a consequence of producing a higher $Q^2$ system. In the upper region of Figure~\ref{fig:Kin}, the gluon nPDF is shown as a function of $x$ for values of $Q = 3, 5$~GeV which approximately correspond to the selected $p_T$ regions of the lower plot. 
In addition to that of EPS09 (which is adopted in this work), it is also informative to consider the shape and magnitude of nPDFs provided by other groups. In the same Figure, the DSSZ~\cite{deFlorian:2011fp} gluon nPDF modification at $Q = 3$~GeV is also shown for the region $x > 10^{-4}$. The DSSZ central value is qualitatively similar to EPS09, in the sense that  (anti-)shadowing is expected at low(moderate)-$x$ values, with the main exception that the magnitude of the nPDF corrections are substantially smaller. This behaviour is not supported by the forward $J/\psi$ data~\cite{Abelev:2013yxa,Aaij:2013zxa}, however, the main point is that $D$ measurements at LHCb are sensitive to both the low-$x$ ($x < 10^{-2}$) shadowing regime, as well as the anti-shadowing regime for values of $x \simeq 0.05$, and will therefore have large impact of future global analyses. It is also worth pointing out that the recent nCTEQ15 analysis~\cite{Kovarik:2015cma} prefers a stronger modification of the gluon PDF at low-$x$ values than that of EPS09.
The sensitivity of high-$p_T$ heavy quark measurements at LHCb to the gluon PDF at large-$x$ has been discussed here~\cite{Cacciari:2015fta}.

\begin{figure}[ht!]
\begin{center}
\makebox{\includegraphics[width=1.1\columnwidth]{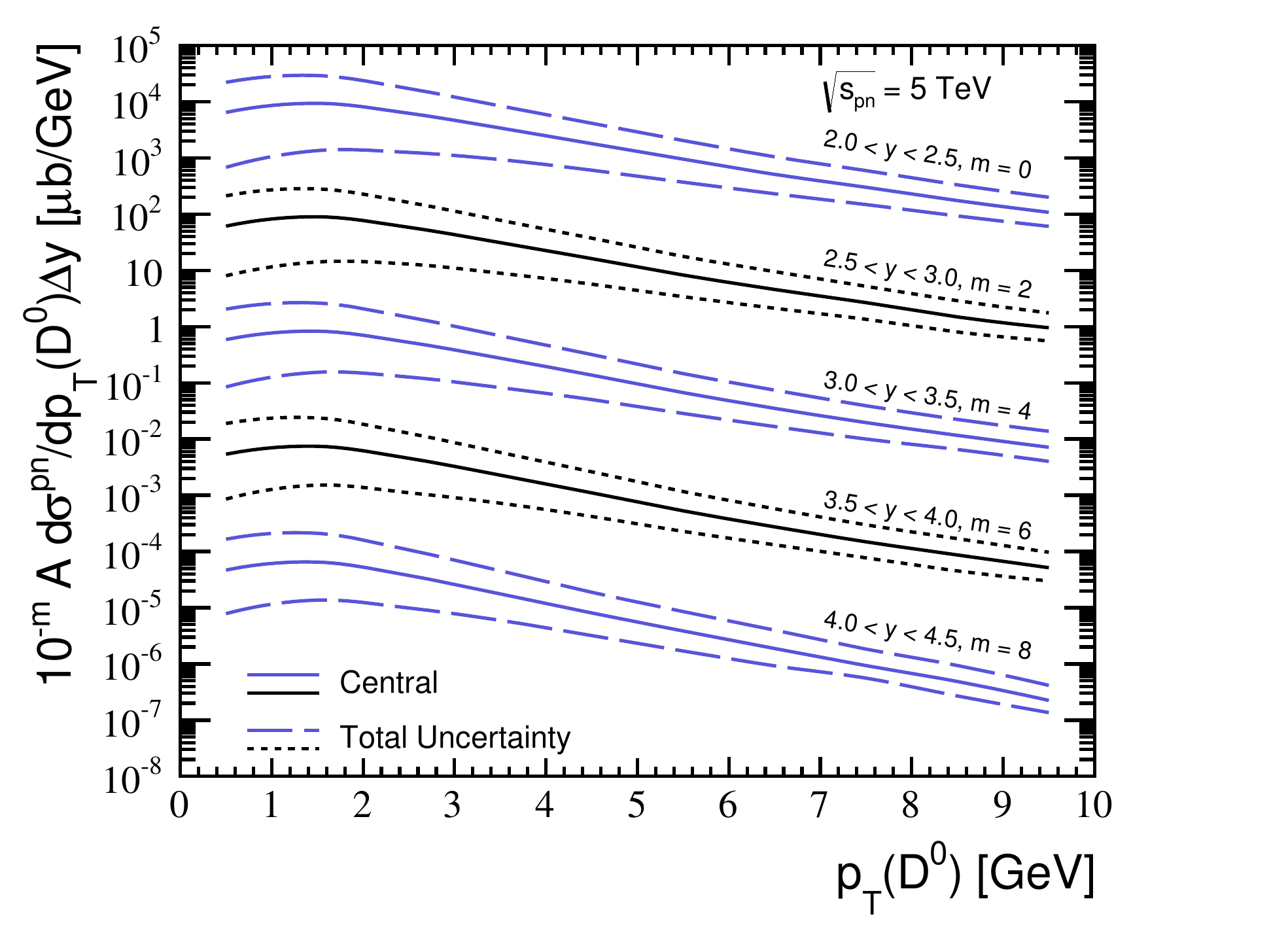}}
\end{center}
\vspace{-8mm}
\caption{Differential $D^0$ cross section in forward $p\rm Pb$ collisions at 5~TeV. Both the central value and total uncertainty for each rapidity bin is shown, and a multiplication factor of $10^{-m}$ has been applied to different rapidity bins as indicated on the plot.}
\label{fig:dsig}
\end{figure}

In the following, predictions are provided for $D^0$ differential cross sections in the Lab frame in forward collisions at 5~TeV. The fragmentation fraction $f(c\rightarrow D^0) = 0.565$~\cite{Aaij:2013mga} is applied. Although only $D^0$ predictions are presented, it is reasonable to apply a ratio of fragmentation fractions to obtain $D_s$, and $D^{\pm}$ predictions, as differences in the distribution of these different flavours due to fragmentation is negligible as compared to scale uncertainties. The results are presented in Figure~\ref{fig:dsig}, where the $p_T$ distributions for different rapidity bins are shown separately. Note that to visually distinguish these different distributions, a rapidity dependent multiplication factor of $10^{-m}$ has been applied. 

The uncertainties at low-$p_T$ are substantial, predominantly due to large scale uncertainties arising at low $Q^2$, but also due to sizeable (n)PDF uncertainties. Consequently, direct constraints on the size of CNM effects in $D$ production are unlikely to come from differential cross section measurements due to the dominance of scale uncertainties. Although only predictions for forward collisions are provided here at 5~TeV, double differential predictions including central values and total uncertainties for forward and backward collisions at 5 and 8~TeV are provided in the Appendix.

\section{Ratio predictions}
A more suitable observable for assessing the size of CNM effects is $R_{fb}$~(\ref{Rfb}) ---  the ratio of cross sections in forward and backward collisions performed in the COM frame. This observable has the benefit that theoretical uncertainties partially cancel in this ratio, with the exception of the nPDF uncertainties. The scale and charm mass uncertainties do not cancel exactly, since varying these parameters results in an altered sampling of the nPDFs. Predictions of $R_{fb}$ will be presented differentially in $p_T$ and $y$ of the $D$ hadron\footnote{Given the high statistics expected in the available data set, measurements of $R_{fb}$ performed double differentially in $p_T$ and $y$ may be feasible. Corresponding predictions are available on request.}.
When theoretical uncertainties are assessed for $R_{fb}$, the variation of parameters is computed simultaneously for forward and backward differential cross sections.

\begin{figure}[ht!]
\begin{center}
\makebox{\includegraphics[width=1.1\columnwidth]{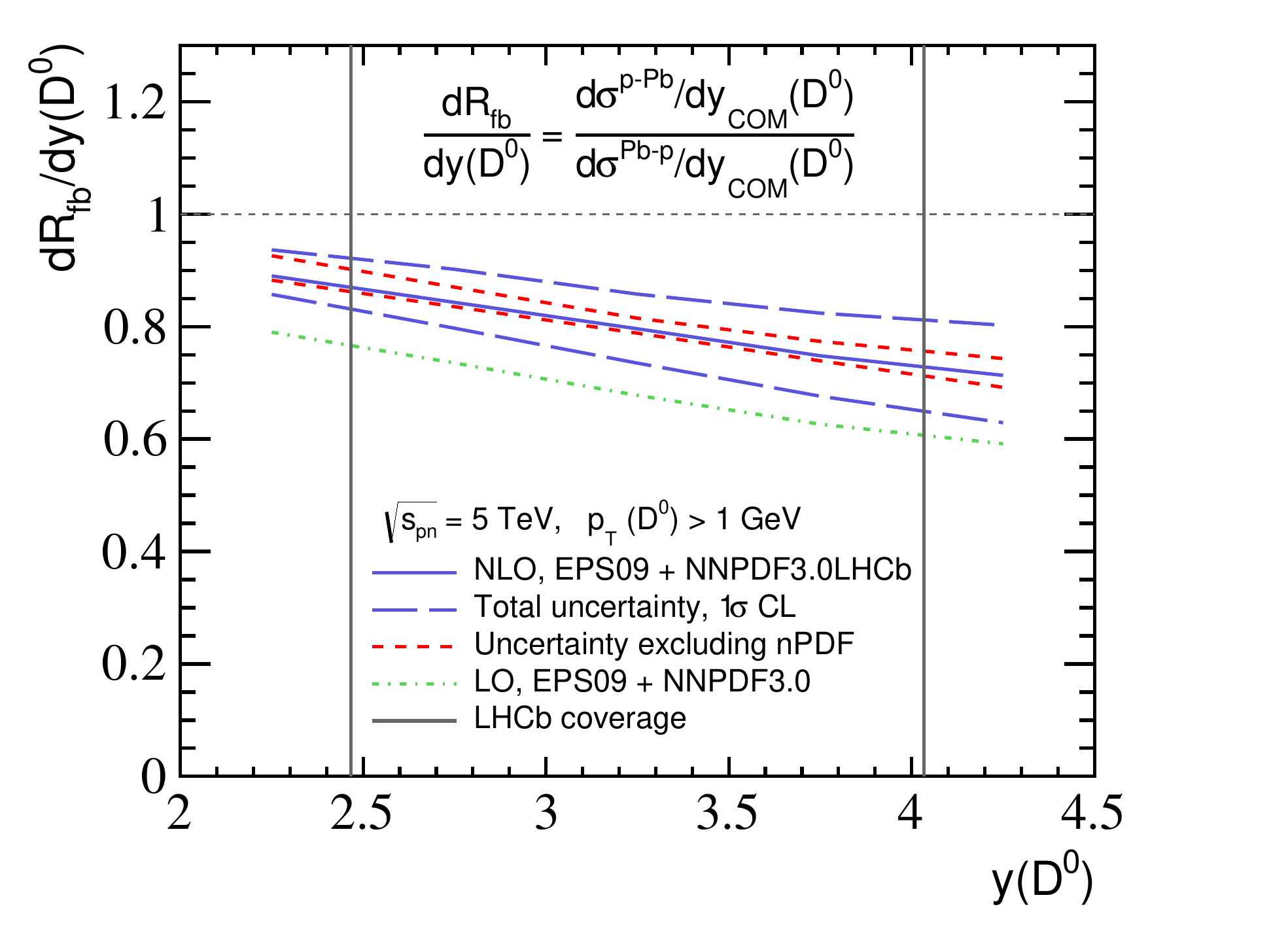}}
\end{center}
\vspace{-8mm}
\caption{Predictions of $R_{fb}$ in $p\rm Pb$ collisions with respect to rapidity for $D^0$ production at 5~TeV. Both the total uncertainty, and uncertainty excluding the dominant nPDF uncertainties are shown. The experimentally accessible region to LHCb is also highlighted.}
\label{fig:Rfbdy}
\end{figure}

In Figure~\ref{fig:Rfbdy}, differential $R_{fb}$ predictions for $D^0$ hadrons with respect to rapidity at 5~TeV are provided. Both the total uncertainty, and uncertainty excluding the nPDF uncertainties are shown demonstrating that nPDFs uncertainties are entirely dominant. Across the accessible kinematic region of $2.47 < y < 4.03$ (limited by the overlap of forward and backward collisions), the central NLO EPS09 value ranges from 0.85 at low rapidity too 0.7 at high rapidity. At higher rapidity, the low-$x$ region of the nPDFs is sampled more frequently in forward collisions, and consequently this ratio is reduced due to shadowing. The LO prediction is systematically lower, primarily due to the fact that the NLO kinematics smear the nPDF sampling at low-$x$ to larger average values where the shadowing effect is less emphasised.

\begin{figure}[ht!]
\begin{center}
\makebox{\includegraphics[width=1.1\columnwidth]{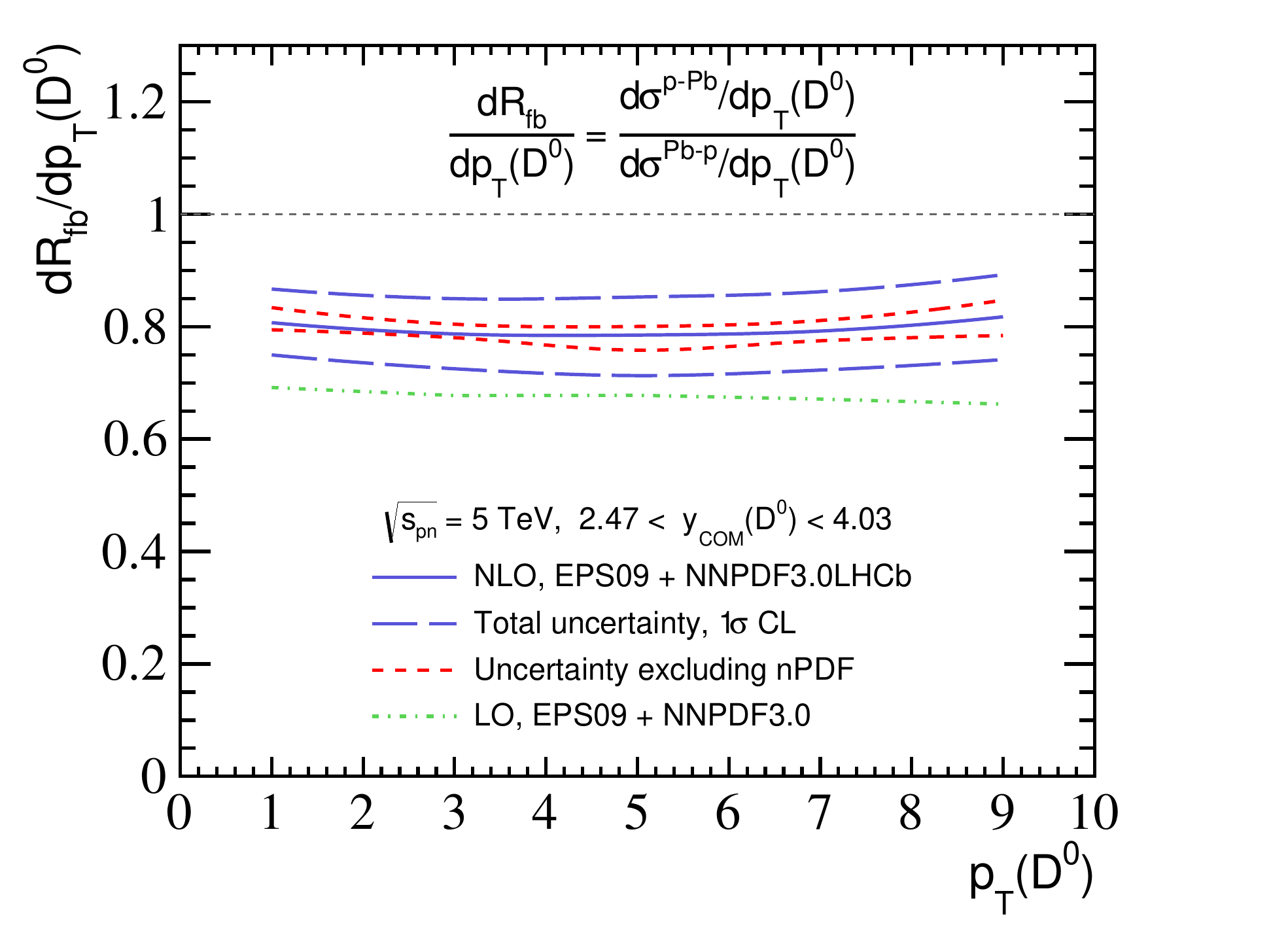}}
\end{center}
\vspace{-8mm}
\caption{The same as Figure~\ref{fig:Rfbdy}, but differentially in $p_T$. This prediction is also for $p\rm Pb$ collisions at 5~TeV.}
\label{fig:Rfbdpt}
\end{figure}

In Figure~\ref{fig:Rfbdpt}, similar predictions are provided differentially in $p_T$. The nPDF uncertainties are entirely dominant in this case too, however the central value is observed to be approximately flat with respect to $p_T$. This behaviour can be understood by re-examining Figure~\ref{fig:Kin}, where the sampled region of the nPDFs for different $p_T$ regions is shown. As the $p_T$ cut of the $D^0$ is increased, the sampled values of $x_1$, and $x_2$ increase. This partially reduces the shadowing effect in forward collisions, but simultaneously increases the anti-shadowing effect in backward collisions. These two effects approximately compensate one another in the considered phase space region. The numerical values of these predictions are provided in the Appendix.

\section{Conclusions}
Predictions for double differential cross sections and forward-backward ratios of $D$ production in $p\rm Pb$ collisions have been provided. These predictions demonstrate that $D$ measurements at LHCb are sensitive to CNM effects. In particular, it is shown that isolating specific $p_T$ regions within the LHCb acceptance can provide sensitivity to both shadowing and anti-shadowing. 

It has to be emphasised that the low-$x$ region of nPDFs is currently not well constrained by data, and consequently different nPDF parameterisations can result in conflicting predictions for the size CNM effects in the forward region. It is essential to perform measurements which extend the current constraints in the region $x < 10^{-3}$. In addition to the proposed $D$ measurements, low mass Drell-Yan measurements should also be performed~\cite{Armesto:2013kqa,Arleo:2015qiv}.

\section{Acknowledgements}
I acknowledge previous collaboration with Uli Haisch, Ben Pecjak, Emanuele Re, Juan Rojo, Luca Rottoli, and Jim Talbert which was relevant for the current work. I'm particularly grateful for discussions and comments received from Emanuele, Juan, and Graeme Watt. I'd also like to thank Patrick Spradlin and Katharina M\"uller for useful comments. Finally, thanks to James Currie for sound advice in times of crisis.

\bibliography{HVQ}

\newpage
\section{Appendix} 
\label{App}
The double differential cross sections for forward and backward $D^0$ production in the Laboratory frame at 5 and 8~TeV are provided in the following Tables. The NLO predictions for $R_{fb}$ provided in Figure~\ref{fig:Rfbdy} and Figure~\ref{fig:Rfbdpt} are also provided, and in addition those at 8~TeV. As previously mentioned, double differential predictions of $R_{fb}$ are available upon request.

\renewcommand*{\arraystretch}{1.4}
\begin{table}[ht!]
  \centering
  \scriptsize
  \begin{tabular}{@{} c|rcl|rcl|rcl|rcl|rcl@{}}
      \hline
   \multicolumn{16}{c}{$A\frac{d^2\sigma^{pn}(D^0)(y,p_T)}{dy^Ddp_T^D} \Delta y \,~ (mb/{\rm GeV})$}  \\ \hline\hline
      	$p_T^D$~(GeV) & \multicolumn{15}{c}{$y^D$}  \\ \hline
	 & \multicolumn{3}{c|}{$2.0-2.5$} &  \multicolumn{3}{c|}{$2.5-3.0$}
	& \multicolumn{3}{c|}{$3.0-3.5$}	 &  \multicolumn{3}{c|}{$3.5-4.0$}	& \multicolumn{3}{c}{$4.0-4.5$} \\ \hline
$0.0-1.0 $ \hspace{-0.2cm} &$6.43$ & \hspace{-0.2cm} & $^{+15.61}_{-5.75}$ &$6.11$ & \hspace{-0.2cm} & $^{+15.04}_{-5.31}$ &$5.86$ & \hspace{-0.2cm} & $^{+14.55}_{-5.02}$ &$5.37$ & \hspace{-0.2cm} & $^{+13.51}_{-4.51}$ &$4.67$ & \hspace{-0.2cm} & $^{+11.87}_{-3.89}$ \\
$1.0-2.0 $ \hspace{-0.2cm} &$9.33$ & \hspace{-0.2cm} & $^{+19.76}_{-7.99}$ &$8.95$ & \hspace{-0.2cm} & $^{+19.08}_{-7.55}$ &$8.27$ & \hspace{-0.2cm} & $^{+17.93}_{-6.73}$ &$7.43$ & \hspace{-0.2cm} & $^{+16.29}_{-5.91}$ &$6.48$ & \hspace{-0.2cm} & $^{+14.58}_{-5.11}$ \\
$2.0-3.0 $ \hspace{-0.2cm} &$6.2$ & \hspace{-0.2cm} & $^{+10.85}_{-4.95}$ &$5.81$ & \hspace{-0.2cm} & $^{+10.29}_{-4.52}$ &$5.24$ & \hspace{-0.2cm} & $^{+9.41}_{-3.98}$ &$4.54$ & \hspace{-0.2cm} & $^{+8.11}_{-3.43}$ &$3.74$ & \hspace{-0.2cm} & $^{+6.92}_{-2.76}$ \\
$3.0-4.0 $ \hspace{-0.2cm} &$3.39$ & \hspace{-0.2cm} & $^{+4.98}_{-2.46}$ &$3.11$ & \hspace{-0.2cm} & $^{+4.65}_{-2.22}$ &$2.74$ & \hspace{-0.2cm} & $^{+4.12}_{-1.92}$ &$2.28$ & \hspace{-0.2cm} & $^{+3.49}_{-1.55}$ &$1.76$ & \hspace{-0.2cm} & $^{+2.77}_{-1.16}$ \\
$4.0-5.0 $ \hspace{-0.2cm} &$1.79$ & \hspace{-0.2cm} & $^{+2.33}_{-1.19}$ &$1.61$ & \hspace{-0.2cm} & $^{+2.12}_{-1.05}$ &$1.37$ & \hspace{-0.2cm} & $^{+1.81}_{-0.87}$ &$1.1$ & \hspace{-0.2cm} & $^{+1.51}_{-0.68}$ &$0.81$ & \hspace{-0.2cm} & $^{+1.07}_{-0.49}$ \\
$5.0-6.0 $ \hspace{-0.2cm} &$0.95$ & \hspace{-0.2cm} & $^{+1.08}_{-0.58}$ &$0.82$ & \hspace{-0.2cm} & $^{+0.94}_{-0.48}$ &$0.67$ & \hspace{-0.2cm} & $^{+0.79}_{-0.38}$ &$0.53$ & \hspace{-0.2cm} & $^{+0.64}_{-0.3}$ &$0.39$ & \hspace{-0.2cm} & $^{+0.47}_{-0.22}$ \\
$6.0-7.0 $ \hspace{-0.2cm} &$0.51$ & \hspace{-0.2cm} & $^{+0.54}_{-0.28}$ &$0.46$ & \hspace{-0.2cm} & $^{+0.49}_{-0.25}$ &$0.35$ & \hspace{-0.2cm} & $^{+0.39}_{-0.18}$ &$0.27$ & \hspace{-0.2cm} & $^{+0.3}_{-0.14}$ &$0.19$ & \hspace{-0.2cm} & $^{+0.21}_{-0.1}$ \\
$7.0-8.0 $ \hspace{-0.2cm} &$0.3$ & \hspace{-0.2cm} & $^{+0.29}_{-0.15}$ &$0.27$ & \hspace{-0.2cm} & $^{+0.25}_{-0.13}$ &$0.2$ & \hspace{-0.2cm} & $^{+0.19}_{-0.1}$ &$0.15$ & \hspace{-0.2cm} & $^{+0.15}_{-0.07}$ &$0.09$ & \hspace{-0.2cm} & $^{+0.09}_{-0.04}$ \\
$8.0-9.0 $ \hspace{-0.2cm} &$0.17$ & \hspace{-0.2cm} & $^{+0.16}_{-0.08}$ &$0.15$ & \hspace{-0.2cm} & $^{+0.14}_{-0.07}$ &$0.12$ & \hspace{-0.2cm} & $^{+0.11}_{-0.05}$ &$0.09$ & \hspace{-0.2cm} & $^{+0.08}_{-0.04}$ &$0.05$ & \hspace{-0.2cm} & $^{+0.05}_{-0.02}$ \\
$9.0-10.0 $ \hspace{-0.2cm} &$0.11$ & \hspace{-0.2cm} & $^{+0.09}_{-0.05}$ &$0.1$ & \hspace{-0.2cm} & $^{+0.08}_{-0.04}$ &$0.07$ & \hspace{-0.2cm} & $^{+0.07}_{-0.03}$ &$0.05$ & \hspace{-0.2cm} & $^{+0.05}_{-0.02}$ &$0.02$ & \hspace{-0.2cm} & $^{+0.02}_{-0.01}$
     \end{tabular}
  \caption{\small Predictions for the differential cross-sections for
    $D^0$ meson production in forward $p\rm Pb$ collisions at 5~TeV.}
  \label{5TeVfor}
\end{table}

\begin{table}[ht!]
  \centering
  \scriptsize
  \begin{tabular}{@{} c|rcl|rcl|rcl|rcl|rcl@{}}
      \hline
   \multicolumn{16}{c}{$A\frac{d^2\sigma^{np}(D^0)(y,p_T)}{dy^Ddp_T^D} \Delta y \,~ (mb/{\rm GeV})$}  \\ \hline\hline
      	$p_T^D$~(GeV) & \multicolumn{15}{c}{$y^D$}  \\ \hline
	 & \multicolumn{3}{c|}{$2.0-2.5$} &  \multicolumn{3}{c|}{$2.5-3.0$}
	& \multicolumn{3}{c|}{$3.0-3.5$}	 &  \multicolumn{3}{c|}{$3.5-4.0$}	& \multicolumn{3}{c}{$4.0-4.5$} \\ \hline
$0.0-1.0 $ \hspace{-0.2cm} &$6.86$ & \hspace{-0.2cm} & $^{+16.56}_{-6.16}$ &$6.83$ & \hspace{-0.2cm} & $^{+16.69}_{-6.07}$ &$6.74$ & \hspace{-0.2cm} & $^{+16.73}_{-5.87}$ &$6.59$ & \hspace{-0.2cm} & $^{+16.71}_{-5.6}$ &$6.25$ & \hspace{-0.2cm} & $^{+16.11}_{-5.31}$ \\
$1.0-2.0 $ \hspace{-0.2cm} &$10.02$ & \hspace{-0.2cm} & $^{+21.0}_{-8.66}$ &$9.91$ & \hspace{-0.2cm} & $^{+21.08}_{-8.36}$ &$9.82$ & \hspace{-0.2cm} & $^{+21.17}_{-8.13}$ &$9.33$ & \hspace{-0.2cm} & $^{+20.49}_{-7.58}$ &$8.63$ & \hspace{-0.2cm} & $^{+19.44}_{-6.87}$ \\
$2.0-3.0 $ \hspace{-0.2cm} &$6.7$ & \hspace{-0.2cm} & $^{+11.54}_{-5.41}$ &$6.58$ & \hspace{-0.2cm} & $^{+11.55}_{-5.17}$ &$6.3$ & \hspace{-0.2cm} & $^{+11.21}_{-4.83}$ &$5.83$ & \hspace{-0.2cm} & $^{+10.6}_{-4.37}$ &$5.09$ & \hspace{-0.2cm} & $^{+9.48}_{-3.76}$ \\
$3.0-4.0 $ \hspace{-0.2cm} &$3.77$ & \hspace{-0.2cm} & $^{+5.58}_{-2.77}$ &$3.55$ & \hspace{-0.2cm} & $^{+5.29}_{-2.54}$ &$3.32$ & \hspace{-0.2cm} & $^{+4.96}_{-2.32}$ &$2.97$ & \hspace{-0.2cm} & $^{+4.59}_{-2.01}$ &$2.43$ & \hspace{-0.2cm} & $^{+3.9}_{-1.64}$ \\
$4.0-5.0 $ \hspace{-0.2cm} &$1.96$ & \hspace{-0.2cm} & $^{+2.48}_{-1.3}$ &$1.86$ & \hspace{-0.2cm} & $^{+2.4}_{-1.21}$ &$1.68$ & \hspace{-0.2cm} & $^{+2.19}_{-1.05}$ &$1.42$ & \hspace{-0.2cm} & $^{+1.9}_{-0.87}$ &$1.12$ & \hspace{-0.2cm} & $^{+1.53}_{-0.67}$ \\
$5.0-6.0 $ \hspace{-0.2cm} &$1.06$ & \hspace{-0.2cm} & $^{+1.21}_{-0.64}$ &$0.98$ & \hspace{-0.2cm} & $^{+1.13}_{-0.58}$ &$0.84$ & \hspace{-0.2cm} & $^{+0.97}_{-0.48}$ &$0.68$ & \hspace{-0.2cm} & $^{+0.81}_{-0.37}$ &$0.49$ & \hspace{-0.2cm} & $^{+0.61}_{-0.26}$ \\
$6.0-7.0 $ \hspace{-0.2cm} &$0.57$ & \hspace{-0.2cm} & $^{+0.6}_{-0.32}$ &$0.53$ & \hspace{-0.2cm} & $^{+0.57}_{-0.28}$ &$0.44$ & \hspace{-0.2cm} & $^{+0.48}_{-0.23}$ &$0.36$ & \hspace{-0.2cm} & $^{+0.39}_{-0.18}$ &$0.24$ & \hspace{-0.2cm} & $^{+0.27}_{-0.12}$ \\
$7.0-8.0 $ \hspace{-0.2cm} &$0.34$ & \hspace{-0.2cm} & $^{+0.31}_{-0.17}$ &$0.31$ & \hspace{-0.2cm} & $^{+0.3}_{-0.15}$ &$0.25$ & \hspace{-0.2cm} & $^{+0.25}_{-0.12}$ &$0.19$ & \hspace{-0.2cm} & $^{+0.19}_{-0.08}$ &$0.11$ & \hspace{-0.2cm} & $^{+0.12}_{-0.05}$ \\
$8.0-9.0 $ \hspace{-0.2cm} &$0.21$ & \hspace{-0.2cm} & $^{+0.19}_{-0.09}$ &$0.18$ & \hspace{-0.2cm} & $^{+0.17}_{-0.08}$ &$0.15$ & \hspace{-0.2cm} & $^{+0.15}_{-0.07}$ &$0.1$ & \hspace{-0.2cm} & $^{+0.1}_{-0.04}$ &$0.06$ & \hspace{-0.2cm} & $^{+0.05}_{-0.02}$ \\
$9.0-10.0 $ \hspace{-0.2cm} &$0.13$ & \hspace{-0.2cm} & $^{+0.11}_{-0.06}$ &$0.11$ & \hspace{-0.2cm} & $^{+0.1}_{-0.05}$ &$0.09$ & \hspace{-0.2cm} & $^{+0.08}_{-0.04}$ &$0.06$ & \hspace{-0.2cm} & $^{+0.06}_{-0.02}$ &$0.03$ & \hspace{-0.2cm} & $^{+0.03}_{-0.01}$
     \end{tabular}
  \caption{\small Predictions for the differential cross-sections for
    $D^0$ meson production in backward $p\rm Pb$ collisions at 5~TeV.}
  \label{5TeVbac}
\end{table}

\begin{table}[ht!]
  \scriptsize
  \begin{tabular}{@{} c|rcl|rcl|rcl|rcl|rcl@{}}
	 & \multicolumn{15}{c}{$dR_{fb}/dx$}  \\ \hline
	$p_T^D$~(GeV) & \multicolumn{3}{c|}{$1.5$} &  \multicolumn{3}{c|}{$3.0$}
	& \multicolumn{3}{c|}{$5.0$}	 &  \multicolumn{3}{c|}{$7.0$}	& \multicolumn{3}{c}{$9.0$} \\ \hline
NLO EPS09 \hspace{-0.1cm} &$0.81$ & \hspace{-0.2cm} & $^{+0.06}_{-0.06}$ &$0.79$ & \hspace{-0.2cm} & $^{+0.06}_{-0.06}$ &$0.78$ & \hspace{-0.2cm} & $^{+0.07}_{-0.07}$ &$0.79$ & \hspace{-0.2cm} & $^{+0.07}_{-0.07}$ &$0.82$ & \hspace{-0.2cm} & $^{+0.08}_{-0.08}$ \\ \hline \hline 	
	 $y$ & \multicolumn{3}{c|}{$2.25$} &  \multicolumn{3}{c|}{$2.75$}
	& \multicolumn{3}{c|}{$3.25$}	 &  \multicolumn{3}{c|}{$3.75$}	& \multicolumn{3}{c}{$4.25$} \\ \hline
NLO EPS09 \hspace{-0.1cm} &$0.89$ & \hspace{-0.2cm} & $^{+0.05}_{-0.03}$ &$0.84$ & \hspace{-0.2cm} & $^{+0.06}_{-0.05}$ &$0.80$ & \hspace{-0.2cm} & $^{+0.06}_{-0.06}$ &$0.75$ & \hspace{-0.2cm} & $^{+0.08}_{-0.07}$ &$0.71$ & \hspace{-0.2cm} & $^{+0.09}_{-0.09}$ 
     \end{tabular}
  \caption{\small Differential predictions for $R_{fb}$ in $p\rm Pb$ collisions at 5~TeV.}
  \label{5TeVRfb}
\end{table}

\renewcommand*{\arraystretch}{1.4}
\begin{table}[ht!]
  \centering
  \scriptsize
  \begin{tabular}{@{} c|rcl|rcl|rcl|rcl|rcl@{}}
      \hline
   \multicolumn{16}{c}{$A\frac{d^2\sigma^{pn}(D^0)(y,p_T)}{dy^Ddp_T^D} \Delta y \,~ (mb/{\rm GeV})$}  \\ \hline\hline
      	$p_T^D$~(GeV) & \multicolumn{15}{c}{$y^D$}  \\ \hline
	 & \multicolumn{3}{c|}{$2.0-2.5$} &  \multicolumn{3}{c|}{$2.5-3.0$}
	& \multicolumn{3}{c|}{$3.0-3.5$}	 &  \multicolumn{3}{c|}{$3.5-4.0$}	& \multicolumn{3}{c}{$4.0-4.5$} \\ \hline
$0.0-1.0 $ \hspace{-0.2cm} &$7.51$ & \hspace{-0.2cm} & $^{+18.08}_{-6.72}$ &$7.39$ & \hspace{-0.2cm} & $^{+17.36}_{-6.29}$ &$7.02$ & \hspace{-0.2cm} & $^{+16.81}_{-5.92}$ &$6.69$ & \hspace{-0.2cm} & $^{+15.61}_{-5.42}$ &$6.07$ & \hspace{-0.2cm} & $^{+14.37}_{-4.89}$ \\
$1.0-2.0 $ \hspace{-0.2cm} &$11.01$ & \hspace{-0.2cm} & $^{+22.57}_{-9.33}$ &$10.73$ & \hspace{-0.2cm} & $^{+21.99}_{-8.75}$ &$10.23$ & \hspace{-0.2cm} & $^{+20.64}_{-8.04}$ &$9.47$ & \hspace{-0.2cm} & $^{+18.83}_{-7.14}$ &$8.43$ & \hspace{-0.2cm} & $^{+17.09}_{-6.32}$ \\
$2.0-3.0 $ \hspace{-0.2cm} &$7.59$ & \hspace{-0.2cm} & $^{+12.63}_{-5.85}$ &$7.16$ & \hspace{-0.2cm} & $^{+12.2}_{-5.4}$ &$6.79$ & \hspace{-0.2cm} & $^{+10.91}_{-4.78}$ &$6.07$ & \hspace{-0.2cm} & $^{+9.95}_{-4.12}$ &$5.29$ & \hspace{-0.2cm} & $^{+8.53}_{-3.48}$ \\
$3.0-4.0 $ \hspace{-0.2cm} &$4.37$ & \hspace{-0.2cm} & $^{+6.12}_{-3.09}$ &$4.12$ & \hspace{-0.2cm} & $^{+5.64}_{-2.76}$ &$3.73$ & \hspace{-0.2cm} & $^{+5.15}_{-2.39}$ &$3.32$ & \hspace{-0.2cm} & $^{+4.33}_{-1.96}$ &$2.71$ & \hspace{-0.2cm} & $^{+3.5}_{-1.53}$ \\
$4.0-5.0 $ \hspace{-0.2cm} &$2.42$ & \hspace{-0.2cm} & $^{+3.04}_{-1.57}$ &$2.29$ & \hspace{-0.2cm} & $^{+2.65}_{-1.38}$ &$2.02$ & \hspace{-0.2cm} & $^{+2.31}_{-1.16}$ &$1.71$ & \hspace{-0.2cm} & $^{+1.88}_{-0.89}$ &$1.36$ & \hspace{-0.2cm} & $^{+1.45}_{-0.68}$ \\
$5.0-6.0 $ \hspace{-0.2cm} &$1.36$ & \hspace{-0.2cm} & $^{+1.45}_{-0.77}$ &$1.24$ & \hspace{-0.2cm} & $^{+1.26}_{-0.66}$ &$1.06$ & \hspace{-0.2cm} & $^{+1.05}_{-0.52}$ &$0.87$ & \hspace{-0.2cm} & $^{+0.86}_{-0.4}$ &$0.68$ & \hspace{-0.2cm} & $^{+0.57}_{-0.29}$ \\
$6.0-7.0 $ \hspace{-0.2cm} &$0.78$ & \hspace{-0.2cm} & $^{+0.74}_{-0.41}$ &$0.7$ & \hspace{-0.2cm} & $^{+0.63}_{-0.34}$ &$0.59$ & \hspace{-0.2cm} & $^{+0.52}_{-0.27}$ &$0.46$ & \hspace{-0.2cm} & $^{+0.39}_{-0.19}$ &$0.36$ & \hspace{-0.2cm} & $^{+0.29}_{-0.13}$ \\
$7.0-8.0 $ \hspace{-0.2cm} &$0.48$ & \hspace{-0.2cm} & $^{+0.41}_{-0.23}$ &$0.42$ & \hspace{-0.2cm} & $^{+0.33}_{-0.18}$ &$0.33$ & \hspace{-0.2cm} & $^{+0.27}_{-0.14}$ &$0.26$ & \hspace{-0.2cm} & $^{+0.21}_{-0.1}$ &$0.2$ & \hspace{-0.2cm} & $^{+0.13}_{-0.06}$ \\
$8.0-9.0 $ \hspace{-0.2cm} &$0.29$ & \hspace{-0.2cm} & $^{+0.24}_{-0.13}$ &$0.26$ & \hspace{-0.2cm} & $^{+0.2}_{-0.1}$ &$0.2$ & \hspace{-0.2cm} & $^{+0.15}_{-0.07}$ &$0.15$ & \hspace{-0.2cm} & $^{+0.11}_{-0.05}$ &$0.11$ & \hspace{-0.2cm} & $^{+0.07}_{-0.04}$ \\
$9.0-10.0 $ \hspace{-0.2cm} &$0.19$ & \hspace{-0.2cm} & $^{+0.15}_{-0.08}$ &$0.17$ & \hspace{-0.2cm} & $^{+0.1}_{-0.06}$ &$0.12$ & \hspace{-0.2cm} & $^{+0.09}_{-0.04}$ &$0.1$ & \hspace{-0.2cm} & $^{+0.07}_{-0.03}$ &$0.06$ & \hspace{-0.2cm} & $^{+0.05}_{-0.02}$
     \end{tabular}
  \caption{\small Predictions for the differential cross-sections for
    $D^0$ meson production in forward $p\rm Pb$ collisions at 8~TeV.}
  \label{8TeVfor}
\end{table}

\renewcommand*{\arraystretch}{1.4}
\begin{table}[ht!]
  \centering
  \scriptsize
  \begin{tabular}{@{} c|rcl|rcl|rcl|rcl|rcl@{}}
      \hline
   \multicolumn{16}{c}{$A\frac{d^2\sigma^{pn}(D^0)(y,p_T)}{dy^Ddp_T^D} \Delta y \,~ (mb/{\rm GeV})$}  \\ \hline\hline
      	$p_T^D$~(GeV) & \multicolumn{15}{c}{$y^D$}  \\ \hline
	 & \multicolumn{3}{c|}{$2.0-2.5$} &  \multicolumn{3}{c|}{$2.5-3.0$}
	& \multicolumn{3}{c|}{$3.0-3.5$}	 &  \multicolumn{3}{c|}{$3.5-4.0$}	& \multicolumn{3}{c}{$4.0-4.5$} \\ \hline
$0.0-1.0 $ \hspace{-0.2cm} &$7.81$ & \hspace{-0.2cm} & $^{+19.09}_{-7.22}$ &$7.87$ & \hspace{-0.2cm} & $^{+19.4}_{-7.08}$ &$7.86$ & \hspace{-0.2cm} & $^{+19.43}_{-6.95}$ &$7.79$ & \hspace{-0.2cm} & $^{+19.51}_{-6.76}$ &$7.57$ & \hspace{-0.2cm} & $^{+19.04}_{-6.49}$ \\
$1.0-2.0 $ \hspace{-0.2cm} &$11.55$ & \hspace{-0.2cm} & $^{+24.21}_{-10.11}$ &$11.55$ & \hspace{-0.2cm} & $^{+24.27}_{-9.9}$ &$11.46$ & \hspace{-0.2cm} & $^{+24.41}_{-9.54}$ &$11.28$ & \hspace{-0.2cm} & $^{+23.73}_{-9.04}$ &$10.69$ & \hspace{-0.2cm} & $^{+22.98}_{-8.47}$ \\
$2.0-3.0 $ \hspace{-0.2cm} &$8.11$ & \hspace{-0.2cm} & $^{+13.53}_{-6.46}$ &$7.86$ & \hspace{-0.2cm} & $^{+13.6}_{-6.2}$ &$7.72$ & \hspace{-0.2cm} & $^{+13.08}_{-5.81}$ &$7.35$ & \hspace{-0.2cm} & $^{+12.44}_{-5.27}$ &$6.76$ & \hspace{-0.2cm} & $^{+11.48}_{-4.65}$ \\
$3.0-4.0 $ \hspace{-0.2cm} &$4.74$ & \hspace{-0.2cm} & $^{+6.64}_{-3.41}$ &$4.53$ & \hspace{-0.2cm} & $^{+6.47}_{-3.19}$ &$4.34$ & \hspace{-0.2cm} & $^{+6.04}_{-2.87}$ &$3.98$ & \hspace{-0.2cm} & $^{+5.61}_{-2.53}$ &$3.58$ & \hspace{-0.2cm} & $^{+4.76}_{-2.09}$ \\
$4.0-5.0 $ \hspace{-0.2cm} &$2.64$ & \hspace{-0.2cm} & $^{+3.29}_{-1.76}$ &$2.55$ & \hspace{-0.2cm} & $^{+3.08}_{-1.59}$ &$2.35$ & \hspace{-0.2cm} & $^{+2.8}_{-1.39}$ &$2.11$ & \hspace{-0.2cm} & $^{+2.47}_{-1.15}$ &$1.77$ & \hspace{-0.2cm} & $^{+1.92}_{-0.88}$ \\
$5.0-6.0 $ \hspace{-0.2cm} &$1.47$ & \hspace{-0.2cm} & $^{+1.59}_{-0.88}$ &$1.4$ & \hspace{-0.2cm} & $^{+1.46}_{-0.77}$ &$1.26$ & \hspace{-0.2cm} & $^{+1.37}_{-0.67}$ &$1.12$ & \hspace{-0.2cm} & $^{+1.09}_{-0.54}$ &$0.9$ & \hspace{-0.2cm} & $^{+0.82}_{-0.39}$ \\
$6.0-7.0 $ \hspace{-0.2cm} &$0.86$ & \hspace{-0.2cm} & $^{+0.86}_{-0.47}$ &$0.81$ & \hspace{-0.2cm} & $^{+0.74}_{-0.4}$ &$0.71$ & \hspace{-0.2cm} & $^{+0.65}_{-0.34}$ &$0.6$ & \hspace{-0.2cm} & $^{+0.54}_{-0.25}$ &$0.46$ & \hspace{-0.2cm} & $^{+0.38}_{-0.17}$ \\
$7.0-8.0 $ \hspace{-0.2cm} &$0.51$ & \hspace{-0.2cm} & $^{+0.46}_{-0.26}$ &$0.49$ & \hspace{-0.2cm} & $^{+0.41}_{-0.22}$ &$0.41$ & \hspace{-0.2cm} & $^{+0.34}_{-0.17}$ &$0.34$ & \hspace{-0.2cm} & $^{+0.27}_{-0.13}$ &$0.25$ & \hspace{-0.2cm} & $^{+0.16}_{-0.08}$ \\
$8.0-9.0 $ \hspace{-0.2cm} &$0.31$ & \hspace{-0.2cm} & $^{+0.27}_{-0.15}$ &$0.3$ & \hspace{-0.2cm} & $^{+0.24}_{-0.13}$ &$0.26$ & \hspace{-0.2cm} & $^{+0.19}_{-0.09}$ &$0.19$ & \hspace{-0.2cm} & $^{+0.15}_{-0.07}$ &$0.14$ & \hspace{-0.2cm} & $^{+0.07}_{-0.04}$ \\
$9.0-10.0 $ \hspace{-0.2cm} &$0.21$ & \hspace{-0.2cm} & $^{+0.17}_{-0.09}$ &$0.19$ & \hspace{-0.2cm} & $^{+0.14}_{-0.08}$ &$0.16$ & \hspace{-0.2cm} & $^{+0.12}_{-0.06}$ &$0.13$ & \hspace{-0.2cm} & $^{+0.07}_{-0.04}$ &$0.08$ & \hspace{-0.2cm} & $^{+0.05}_{-0.02}$
     \end{tabular}
  \caption{\small Predictions for the differential cross-sections for
    $D^0$ meson production in backward $p{\rm Pb}$ collisions at 8~TeV.}
  \label{8TeVbac}
\end{table}


\begin{table}[ht!]
  \scriptsize
  \begin{tabular}{@{} c|rcl|rcl|rcl|rcl|rcl@{}}
	 & \multicolumn{15}{c}{$dR_{fb}/dx$}  \\ \hline
	$p_T^D$~(GeV) & \multicolumn{3}{c|}{$1.5$} &  \multicolumn{3}{c|}{$3.0$}
	& \multicolumn{3}{c|}{$5.0$}	 &  \multicolumn{3}{c|}{$7.0$}	& \multicolumn{3}{c}{$9.0$} \\ \hline
NLO EPS09 \hspace{-0.1cm} &$0.85$ & \hspace{-0.2cm} & $^{+0.06}_{-0.05}$ &$0.83$ & \hspace{-0.2cm} & $^{+0.06}_{-0.05}$ &$0.82$ & \hspace{-0.2cm} & $^{+0.05}_{-0.06}$ &$0.79$ & \hspace{-0.2cm} & $^{+0.06}_{-0.06}$ &$0.78$ & \hspace{-0.2cm} & $^{+0.07}_{-0.07}$ \\ \hline \hline 	
	 $y$ & \multicolumn{3}{c|}{$2.25$} &  \multicolumn{3}{c|}{$2.75$}
	& \multicolumn{3}{c|}{$3.25$}	 &  \multicolumn{3}{c|}{$3.75$}	& \multicolumn{3}{c}{$4.25$} \\ \hline
NLO EPS09 \hspace{-0.1cm} &$0.92$ & \hspace{-0.2cm} & $^{+0.05}_{-0.02}$ &$0.88$ & \hspace{-0.2cm} & $^{+0.07}_{-0.03}$ &$0.84$ & \hspace{-0.2cm} & $^{+0.06}_{-0.05}$ &$0.79$ & \hspace{-0.2cm} & $^{+0.06}_{-0.06}$ &$0.74$ & \hspace{-0.2cm} & $^{+0.07}_{-0.07}$ 
     \end{tabular}
  \caption{\small Differential predictions for $R_{fb}$ in $p\rm Pb$ collisions at 8~TeV.}
  \label{8TeVRfb}
\end{table}

\end{document}